%% file: main.tex
\documentclass[10pt,twocolumn,letterpaper]{article}

\usepackage[letterpaper, top=0.75in, left=0.75in, right=0.75in,
            bottom=1in, columnsep=0.25in]{geometry}
\usepackage{newtxtext,newtxmath}   
\usepackage[utf8]{inputenc}
\usepackage[T1]{fontenc}
\usepackage{amsmath}
\allowdisplaybreaks[4]
\usepackage{graphicx}
\usepackage{booktabs}
\usepackage{multirow}
\usepackage{subcaption}
\usepackage{multicol}
\usepackage{pdflscape}
\usepackage{xcolor}
\usepackage{xspace}
\usepackage[numbers, sort&compress]{natbib}
\usepackage[colorlinks=true, linkcolor=blue!60!black, citecolor=blue!60!black, urlcolor=blue!60!black]{hyperref}

\graphicspath{{figures/}}

\usepackage{microtype}   

\newcommand{\MET}{\ensuremath{E_{\mathrm{T}}^{\mathrm{miss}}}\xspace}
\makeatletter
\renewcommand\section{\@startsection{section}{1}{\z@}%
  {-3.5ex \@plus -1ex \@minus -.2ex}%
  {2.3ex \@plus.2ex}%
  {\raggedright\normalfont\Large\bfseries}}
\renewcommand\subsection{\@startsection{subsection}{2}{\z@}%
  {-3.25ex\@plus -1ex \@minus -.2ex}%
  {1.5ex \@plus .2ex}%
  {\raggedright\normalfont\large\bfseries}}
\renewcommand\subsubsection{\@startsection{subsubsection}{3}{\z@}%
  {-3.25ex\@plus -1ex \@minus -.2ex}%
  {1.5ex \@plus .2ex}%
  {\raggedright\normalfont\normalsize\bfseries}}
\makeatother

\newcommand{\new}[1]{\textcolor{blue}{#1}}

\newcommand{\lqbt}{\ensuremath{\mathrm{LQ} \to \mathrm{b}\tau}}
\newcommand{\afourl}{\ensuremath{A \to 4\ell}}
\newcommand{\htautau}{\ensuremath{h^{0} \to \tau\tau}}
\newcommand{\htaunu}{\ensuremath{h^{\pm} \to \tau\nu}}
\newcommand{\pt}{\ensuremath{p_{\mathrm{T}}}}
\title{\textbf{Distillation of a Normalising Flow into a Compact Symbolic
Expression for Real-Time Anomaly Detection at the LHC Level-1 Trigger}}

\author{%
\parbox[t]{0.95\textwidth}{\centering Justin Bollhalder$^{1,2}$, Florencia Canelli$^{3}$, Valentina Guglielmi$^{3}$, Roope Oskari Niemi$^{4}$,\\ Maurizio Pierini$^{4}$, Chang Sun$^{5}$, Francesco Vaselli$^{6}$}\\[4pt]
{\small $^{1}$Department of Mathematics, University of Zurich, Zurich, Switzerland}\\
{\small $^{2} $Lake Lucerne Institute, Vitznau, Switzerland}\\
{\small $^{3}$Physik-Institut, Universit\"at Z\"urich, Z\"urich, Switzerland}\\
{\small $^{4}$European Organization for Nuclear Research (CERN), CH-1211 Geneva 23, Switzerland}\\
{\small $^{5}$California Institute of Technology, Pasadena, CA, USA}\\
{\small $^{6}$Scuola Normale Superiore, Pisa, Italy, and Istituto Nazionale di Fisica Nucleare (INFN), Pisa, Italy}\\[4pt]
{\small $^{*}$Corresponding authors: \href{mailto:justinroger.bollhalder@uzh.ch}{justinroger.bollhalder@uzh.ch} (J.~Bollhalder), \href{mailto:valentina.guglielmi@physik.uzh.ch}{valentina.guglielmi@physik.uzh.ch} (V.~Guglielmi)}\\
}

\date{}

\begin{document}

\maketitle

\begin{abstract}
At the CERN Large Hadron Collider, anomaly detection techniques are becoming popular alternatives to traditional signature-aware new physics searches. For this approach to avoid any bias induced by the real-time event selection applied in the trigger system of ATLAS and CMS, it should be implemented before the first stage of selection, the Level-1 Trigger. This requires deciding to accept or reject each collision within a few microseconds, a latency budget that excludes the most expressive anomaly-detection models. Knowledge distillation has successfully addressed this issue by replacing a complex anomaly-detection model  with a lighter student network.
This work takes the compression one step further, distilling an anomaly detection model based on a normalising flow into an explicit mathematical formula. Using the SymbolNet symbolic-regression framework with adaptive
pruning, the student is a learned functional form based on a chosen subset of the inputs, providing at once feature selection and compression through knowledge distillation. The final expression contains 577 operations and matches the teacher's discrimination on a test dataset, generalising to four benchmark new physics scenarios. Implemented on a Xilinx Virtex UltraScale+ VU13P FPGA, the symbolic student has a nine-cycle latency, corresponding to 42.8\,ns at the post-route maximum clock frequency, and uses 0.81\% of the device's LUTs and 0.44\% of its DSP slices.
\end{abstract}

\section{Introduction}
\label{sec:intro}

The CERN Large Hadron Collider (LHC) is the ultimate source of scientific data: proton collisions happen in the four large LHC detectors at a nominal rate of 40\,MHz, producing ${\cal O}(1)$ MB of data each time. Such a large volume of raw detector data 
cannot be recorded in full by the ATLAS~\cite{ATLAS:2008xda} and
CMS~\cite{CMS:2008xjf} experiments.
Both experiments
therefore rely on real-time selection algorithms, running in two-stage trigger systems. The trigger determines, in real time,
which collisions are retained for further analysis. In CMS and ATLAS, the first stage
of this selection is the hardware-based Level-1 Trigger
(L1T)~\cite{CMS:2020cmk,CMS:L1TDR}, implemented on Field-Programmable Gate Arrays
(FPGAs). Within a latency budget of a few microseconds, the L1T must reduce
the incoming event rate from ${\cal O}(10)$~MHz to ${\cal O}(100)$~kHz.
The accepted events are passed to the software-based High-Level Trigger (HLT), where more sophisticated reconstruction and selection algorithms can be applied on a more relaxed timescale.

The stringent latency and resource constraints of the L1T have
traditionally favoured algorithms designed to identify specific,
predefined physics signatures. Such targeted selections have been central
to the success of the LHC physics programme, but necessarily rely on
assumptions about how the sought-after phenomena appear in the detector.
This limitation has motivated increasing interest in model-agnostic anomaly
detection~\cite{Kasieczka:2021xcg,Aarrestad:2021oeb}, in which algorithms learn a
description of Standard Model (SM) events and identify collisions that
depart from it. Early studies investigated autoencoder-based methods for
offline and high-level-trigger applications~\cite{Kingma:2013vae,Cerri:2018anp,Knapp:2020adversarial},
while subsequent work demonstrated that machine-learning anomaly detectors
could be implemented directly at the L1T~\cite{Govorkova:2021utb}. This research
programme has led to algorithms such as AXOL1TL and
CICADA~\cite{Gandrakota:2024axoltl} in CMS and GELATO~\cite{Cohen:2025gelato} in
ATLAS. More recently, continuous normalising flows have also
been investigated for real-time anomaly detection at the L1T~\cite{Vaselli:2025fad}.

A central challenge is that the most expressive anomaly-detection models are generally not those most easily deployed under the stringent hardware constraints of the L1T. As proposed in~\cite{pol2023knowledgedistillationanomalydetection}, 
one can use knowledge distillation~\cite{Hinton:2015distill} to train a supervised student to imitate the decision of a much more complex unsupervised teacher. This paradigm has been successfully applied to convolutional autoencoders~\cite{Gerlach:2025brx} and more recently to normalising flows~\cite{companion,Tahseen:2026dnf}. A high-capacity normalising flow, trained exclusively on SM events and using the exact negative
log-likelihood under the learned density as its anomaly score, served as
the teacher for a compact multilayer perceptron (MLP), which served as the student. Ref.~\cite{Tahseen:2026dnf} also investigated the use of a boosted decision tree as a student.
This strategy largely
decouples the expressive model used to learn the SM distribution from the architecture ultimately deployed in the trigger. 

In this work, we retain the teacher--student strategy of
Ref.~\cite{companion} but investigate a fundamentally different
representation for the student. Rather than distilling the normalising-flow
teacher into another neural network, we ask whether its anomaly-detection
behaviour can be represented by a compact \emph{symbolic expression}~\cite{Schmidt:2009science,Cranmer:2020wew}. We use SymbolNet~\cite{Tsoi:2024ypg}, a neural symbolic-regression framework with
adaptive pruning, to learn an explicit function of the trigger-level input
features from the teacher scores. During training, the framework selects
the relevant input features, operations, and terms, yielding a student
composed of arithmetic operations and a limited number of elementary
functions. The objective of this paper is to determine how much of the performance of the normalising-flow
strategy can be retained when the neural-network student of
Ref.~\cite{companion} is replaced by a symbolic representation.

Such a representation offers several potential advantages for real-time
deployment~\cite{Tsoi:2023isc}. The complete trigger algorithm is expressed as an explicit
mathematical formula whose inputs, coefficients, and individual operations
can be inspected directly. Its arithmetic complexity can be determined
before synthesis, while the effects of coefficient and intermediate-value quantisation can be studied operation by operation. Symbolic regression therefore offers a route towards further compression of the distilled model while making the hardware-facing decision rule more transparent.

The main contributions of this work are:
\begin{itemize}
  \item We distil the anomaly scores of the RealNVP teacher from
  Ref.~\cite{companion} into a compact symbolic expression, rather than
  into a conventional machine-learning algorithm;

  \item We quantify how accurately the symbolic student reproduces the
  teacher's anomaly-detection performance across four BSM benchmarks and
  compare it with the neural-network student of Ref.~\cite{companion};

  \item We characterise the symbolic expression from the perspective of
  hardware deployment, reporting its arithmetic complexity, the input
  features it retains, and its FPGA latency and resource usage.
\end{itemize}

The rest of this paper is organised as follows.
Section~\ref{sec:data} describes the data samples.
Section~\ref{sec:teacher} summarises the teacher model taken from
Ref.~\cite{companion}.
Sections~\ref{sec:student}, \ref{sec:target}, and~\ref{sec:training}
present the symbolic student, the distillation target and loss function,
and the training and pruning procedure, respectively.
Section~\ref{sec:performance} compares the anomaly-detection performance
of the symbolic student with the teacher and with the neural-network
student of Ref.~\cite{companion}. Section~\ref{sec:fpga} reports
the FPGA implementation and the corresponding resource comparison, and
Section~\ref{sec:summary} concludes the paper.

\section{Data samples}
\label{sec:data}

We use the same publicly available LHC anomaly-detection benchmark datasets, event selection, and pre-processing as Ref.~\cite{companion}, to which we refer for a detailed description. The samples were originally introduced in Ref.~\cite{Govorkova:2021utb} and released on Zenodo~\cite{Aarrestad:2021v2,zenodo:lq,zenodo:a4l,zenodo:h0tt,zenodo:hctaunu}; their generation and composition are documented in Ref.~\cite{Govorkova:2021dataset}. The same benchmark has also been used in Refs.~\cite{Govorkova:2021utb,Vaselli:2025fad}, allowing the different anomaly-detection approaches to be evaluated on a common dataset.

The SM background consists of simulated proton--proton collision events passing the trigger-like pre-selection described in Ref.~\cite{companion}.
We use the identical data partition as in Ref.~\cite{companion}: 2.8M events for training, 700\,000 for validation, and an independent sample of 2M events for testing.
Performance on anomalous events is evaluated using four beyond-the-Standard-Model (BSM) benchmark samples:
\begin{itemize}
\item $\lqbt$: an $80\,\mathrm{GeV}$ leptoquark, 340\,544 events~\cite{zenodo:lq};
\item $\afourl$: a $50\,\mathrm{GeV}$ neutral scalar, 55\,969 events~\cite{zenodo:a4l};
\item $\htautau$: a $60\,\mathrm{GeV}$ neutral scalar, 691\,283 events~\cite{zenodo:h0tt};
\item $\htaunu$: a $60\,\mathrm{GeV}$ charged scalar, 760\,272 events~\cite{zenodo:hctaunu}.
\end{itemize}
None of the BSM samples is used during training.

Each event is represented by 57 trigger-level input features: transverse
momentum $\pt$, pseudorapidity $\eta$, and azimuthal angle $\phi$ for up
to four muons, four electrons, and ten jets, together with the
missing-transverse-energy information represented by $|\MET|$ and $\phi$,
with its pseudorapidity component set to zero. Missing object entries are
zero-padded to obtain a fixed-size representation, and all features are standardised using statistics determined exclusively from the SM training sample. The same input representation is used for both the teacher and the symbolic student.

\section{The teacher model}
\label{sec:teacher}

The teacher model is the RealNVP normalising flow developed and
validated in Ref.~\cite{companion}, to which we refer for a complete
description of its architecture, training procedure, and robustness studies. Here, we summarise only the aspects required for the distillation procedure used in this work.

The teacher normalising flow learns an invertible mapping between the
57-dimensional input space and a latent space endowed with a standard normal prior. The probability density assigned by the model to each event can therefore be evaluated exactly using the change-of-variables formula. The resulting per-event negative log-likelihood (NLL), $\ell(x) = -\log p(x)$, is used as the teacher anomaly score: events assigned a lower probability by the learned SM density have a larger NLL and are therefore regarded as more anomalous. This construction does not rely on any specific BSM signal hypothesis. In contrast to reconstruction-error-based anomaly scores, and to variational approaches based on a lower bound on the likelihood, a
normalising flow provides an explicit and tractable likelihood under the learned density model.

The RealNVP architecture comprises 8 affine coupling layers with
alternating masks. Each coupling layer contains two independent neural networks that parameterise the scale and translation transformations, respectively; both networks have three hidden layers of 256 units with ReLU activations. The complete model contains approximately 2.6M trainable parameters. It is trained on the 2.8M SM training events by minimising the mean NLL using the Adam optimiser~\cite{Kingma:2014adam}, with an
initial learning rate of $10^{-4}$ and a batch size of 4096. Training is performed for up to 200 epochs, with early stopping after five epochs without improvement in the validation loss and restoration of the best-performing weights. Throughout this study, we use the same trained teacher instance as in Ref.~\cite{companion}.

\begin{figure}[tbp]
  \centering
  \includegraphics[width=\columnwidth]{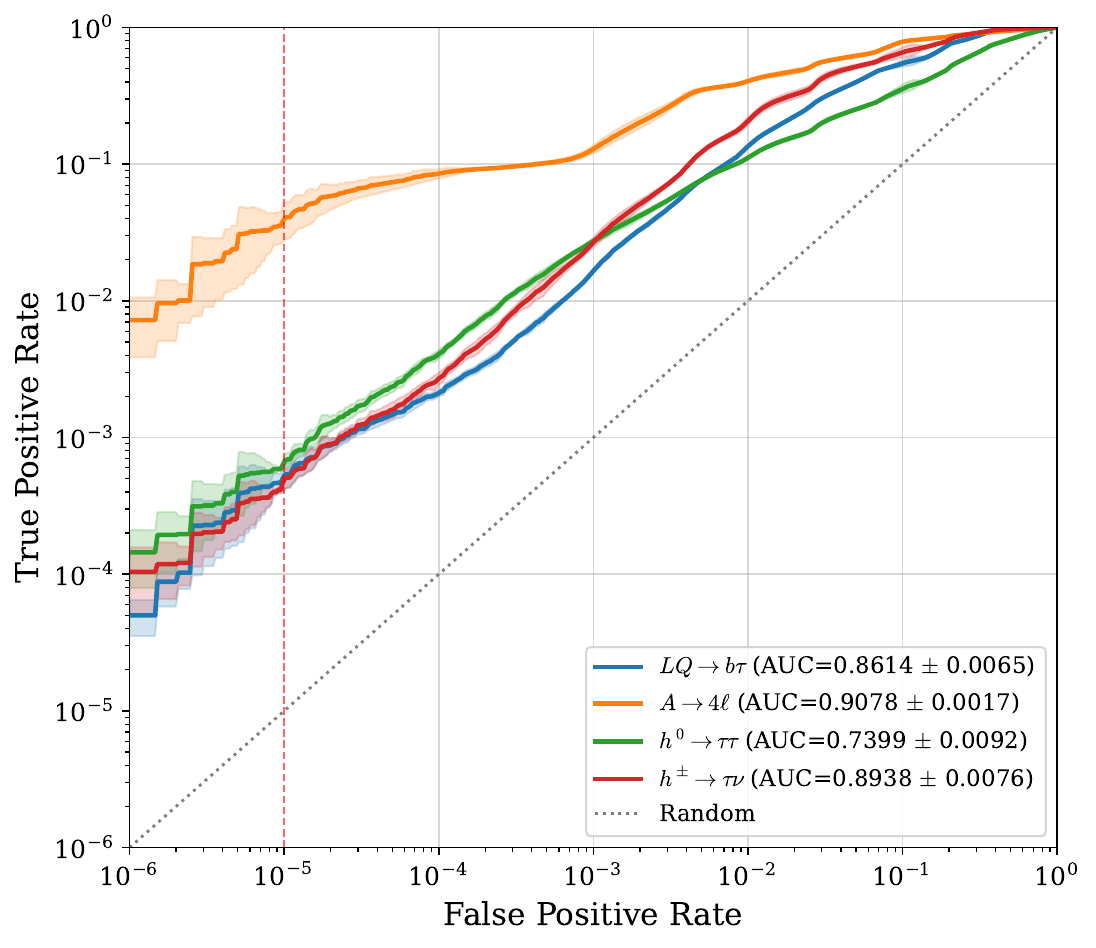}
  \caption{ROC curves of the RealNVP teacher on the four BSM benchmark
  signals, each evaluated against the SM test set. The dotted diagonal
  indicates a random classifier. Figure from Ref.~\cite{companion}.}
  \label{fig:teacher-roc}
\end{figure}

Table~\ref{tab:teacher} summarises the performance of the teacher on the four BSM benchmarks. We report the area under the receiver operating characteristic curve (AUC) and the true-positive rate (TPR) at false-positive rates (FPRs) of $10^{-2}$, $10^{-3}$, $10^{-4}$, and $10^{-5}$. The highest AUC is obtained for the \afourl{} benchmark, whereas the lowest is obtained for \htautau{}, which is more difficult to distinguish from the SM background. The corresponding ROC curves are shown in Figure~\ref{fig:teacher-roc}.

\begin{table}[t]
  \centering
  \small
  \caption{Teacher performance: AUC and TPR at fixed background rates of
  $10^{-2}$, $10^{-3}$, $10^{-4}$ and $10^{-5}$ for the four BSM benchmarks,
  reproducing the values of Ref.~\cite{companion}. All values are given in percent.}
  \label{tab:teacher}
  \setlength{\tabcolsep}{3.5pt}
  \begin{tabular}{lccccc}
    \toprule
    Signal & AUC & \multicolumn{4}{c}{TPR at fixed FPR} \\
    \cmidrule(lr){3-6}
          &        & $10^{-2}$ & $10^{-3}$ & $10^{-4}$ & $10^{-5}$ \\
    \midrule
    \lqbt    & 86.14 & 13.52 & 1.66 & 0.21 & 0.054 \\
    \afourl  & 90.78 & 40.52 & 12.91 & 8.54 & 4.067 \\
    \htautau & 73.99 & 11.13 & 2.75 & 0.41 & 0.068 \\
    \htaunu  & 89.38 & 20.66 & 2.70 & 0.27 & 0.051 \\
    \bottomrule
  \end{tabular}
\end{table}

\section[Symbolic-student architecture]{Symbolic-student architecture}
\label{sec:student}

The symbolic student is trained to reproduce the teacher scores using SymbolNet~\cite{Tsoi:2024ypg}. 
Sparsity is learned jointly during training through trainable pruning thresholds, which determine which input features, weights, and symbolic terms remain in the final expression.
The student consists of three ingredients:
\begin{itemize}
    \item {\bf Input gate:} Each of the 57 input features carries a learnable threshold in $[0,1]$; a feature survives only while its (fixed, unit) gate weight exceeds the threshold. Features whose gates close will not contribute to the final mathematical expression.
    \item {\bf Symbolic layers:} Each hidden layer first computes masked affine combinations of its inputs. Unary slots apply functions assigned from a fixed set of available functions to individual combinations, whereas each product slot multiplies a pair of combinations to generate cross-terms. Every weight and bias has an independent non-negative threshold and remains active only while its magnitude exceeds that threshold: a weight $w$ contributes to the network as $w\,\Theta(|w|-t_w)$, where $\Theta$ is the Heaviside step function ($\Theta(x)=1$ for $x>0$ and $0$ otherwise) and $t_w$ is its learnable threshold, following the dynamic-sparse-training masking scheme of Ref.~\cite{liu2020dynamic} as adapted in SymbolNet. Operator slots have separate gates: closing a unary gate replaces its assigned function with the identity, while closing a binary gate replaces its operator with addition. The gates are evaluated by a hard step function in the forward pass, with the derivative of a sigmoid used as a smooth surrogate during backpropagation.  In this paper, the binary thresholds are frozen, so all eight product slots in each hidden layer remain active.
    \item {\bf Self-pruning through loss:} During the adaptive pruning steps, sparsity regularisers drive the input gates, affine coefficients, and unary operator gates towards target sparsity fractions $\alpha$. The learned thresholds determine which entries remain active during these steps. A subsequent deterministic saliency-ranked cut enforces the final operation budget, after which the mask structure is frozen and the surviving coefficients are re-tuned. Binary gates remain open throughout.
\end{itemize}

The symbolic student has two hidden symbolic layers, each with 32 unary
slots and eight product slots. Its architecture before and after pruning is
summarised in Table~\ref{tab:models}.

\begin{table}[tbp]
  \centering
  \small
  \caption{Symbolic student: the network as configured before training (upper
  block) and the closed-form expression obtained after pruning and
  size-targeted compression (lower block). The slot counts and the function
  set are those made \emph{available} to the optimiser; pruning selects a
  subset of them.}
  \label{tab:models}
  \begin{tabular}{ll}
    \toprule
    Property & Symbolic student \\
    \midrule
    \multicolumn{2}{l}{\emph{As configured (before training)}} \\
    hidden symbolic layers & 2 \\
    slots per layer (unary + product) & $32 + 8$ \\
    available unary functions & $\{\mathrm{id},x^2,\tanh,e^{-x^2},\cos\}$ \\
    available binary operation & $\times$ \\
    inputs offered & 57 \\
    \midrule
    \multicolumn{2}{l}{\emph{After training and compression}} \\
    expression size & 577 CSE operations \\
    inputs read by the expression & 30 of 57 \\
    \bottomrule
  \end{tabular}
\end{table}

The operator set was chosen from exploratory runs with a broader collection
of numerically stable functions. After pruning, $\tanh$ and $e^{-x^2}$ were
consistently the most useful nonlinearities, while $\cos$ was retained less
frequently; the remaining candidates provided little improvement after
compression.
The square is evaluated as $\operatorname{clip}(x,-10,10)^2$, a
numerical guard of the SymbolNet implementation that is kept in the
deployed expression, where it acts on the most anomalous events.

Expression size is measured with SymPy~\cite{Meurer:2017yhf} \texttt{count\_ops} after
common-subexpression elimination (CSE), with each elementary-function call
counted as one operation. The final expression contains 577 operations and
uses 30 input features.

\section[Distillation target and loss function]{Distillation target and loss function}
\label{sec:target}
For ranking-based anomaly detection, the relevant information in the
teacher NLL is its position relative to the SM
distribution, rather than its absolute numerical scale. Following
Ref.~\cite{companion}, we therefore express each event in terms of its rank
among the $N=2.8$M SM training events.

\begin{figure}[tbp]
  \centering
  \includegraphics[width=\columnwidth]{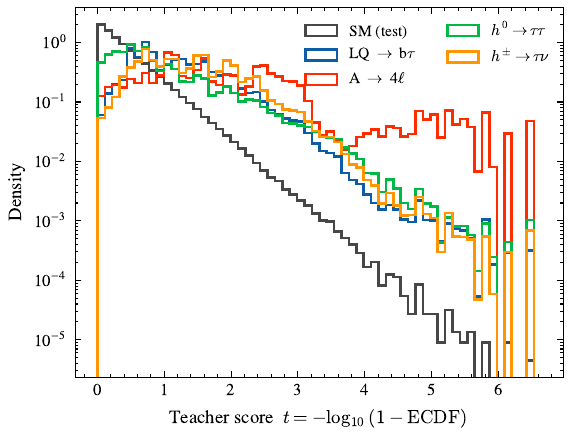}
  \caption{Teacher score distributions after the
  $t=-\log_{10}(1-\mathrm{ECDF})$ transformation for the SM test sample
  and the four anomaly benchmarks. The SM distribution decreases by one
  decade per unit of $t$ by construction, while all four signals extend
  into the high-$t$ tail.}
  \label{fig:teacher-scores}
\end{figure}

Let $\{x_i\}_{i=1}^{N}$ denote the SM training sample and $\ell(x)$ the
teacher NLL of an event. Each event is assigned the empirical cumulative
distribution function (ECDF) value of its NLL over the SM training sample,
and the upper-tail probability is mapped to a logarithmic scale. We use the
same ECDF construction as Ref.~\cite{companion} and define the tail score
\begin{equation}
t(x)
=
-\log_{10}\left[
1-
\frac{
\min\!\left(
\left|\left\{i:\ell(x_i)<\ell(x)\right\}\right|,
N-1
\right)+1
}{
N+1
}
\right].
\label{eq:ecdf}
\end{equation}
The clipping of the empirical rank at $N-1$ prevents the argument of the
logarithm from vanishing for events with an NLL larger than all values in
the training sample, and bounds the score:
\begin{equation}
  0<t(x)\le\log_{10}(N+1)\approx6.45
  \qquad\text{for } N=2.8\times10^{6}.
  \label{eq:tail}
\end{equation}
The logarithmic map matters because the region relevant to a low-rate
trigger is compressed into a narrow interval of ECDF values near one: SM
tail probabilities of $10^{-3}$ and $10^{-5}$ correspond to ECDF values of
$0.999$ and $0.99999$, respectively, whereas one unit of $t$ corresponds to
one decade of SM tail probability.
The four BSM benchmark samples exhibit enhanced populations in the
high-$t$ region, as shown in Figure~\ref{fig:teacher-scores}.

The score $t$ is a non-decreasing function of the teacher NLL and
therefore preserves its ordering, apart from ties introduced by the finite
ECDF resolution and the upper-rank cap. In the numerical evaluation, the
AUC obtained from $t$ agrees with that obtained directly from the NLL to
better than $10^{-7}$.
\begin{table*}[t]
  \centering
  \footnotesize
  \caption{Training steps and hyperparameters for the symbolic student.
  $\alpha = (\alpha_{\mathrm{in}}, \alpha_{\mathrm{w}},
  \alpha_{\mathrm{un}})$ are the sparsity targets for input features,
  weights, and unary operators; products are never pruned. The retuning
  steps keep the checkpoint with the lowest SM validation loss
  (Section~\ref{sec:training}).}
  \label{tab:hyper}
  \begin{tabular}{cll}
    \toprule
    Step & Stage & Symbolic student \\
    \midrule
    1 & plain training & $\beta{=}1$; $100$ ep @ $1.2{\cdot}10^{-3}$ + $60$ @ $4{\cdot}10^{-4}$
      + $30$ @ $1.2{\cdot}10^{-4}$; bs 2048 \\
    2 & light pruning & $\beta{=}1$; $\alpha=(.35,.85,.50)$; 70 ep; lr $10^{-3}$;
      saliency guard (top 12) \\
    3 & stronger pruning & $\beta{=}1$; $\alpha=(.40,.93,.60)$; 55 ep; lr $8{\cdot}10^{-4}$;
      saliency guard (top 12) \\
    4 & reduction to size & $\beta{=}2$; 7--15\,\% saliency-ranked cuts with
      6-ep recoveries; CSE $\le626$ \\
    5 & fixed-structure retuning & $\beta{=}3$; ${\le}90$ ep;
      lr $2{\cdot}10^{-4}$; bs 4096 \\
    6 & low-lr retuning & $\beta{=}4$\new{;} ${\le}60$ ep\new{;}
      lr $6{\cdot}10^{-5}$\new{;} bs 4096 \\
    \bottomrule
  \end{tabular}
\end{table*}


The student is trained to regress the target $y(x)=t(x)$ of
Eq.~(\ref{eq:ecdf}) with a scalar output $z(x)$, which is used directly as
the student anomaly score.
The regression component of the loss is a tail-weighted Huber loss:
\begin{equation}
  \begin{gathered}
    \mathcal{L}_{\mathrm{reg}} =
    \Big\langle
      \bigl(1+\beta\,p(y)^2\bigr)\,
      10^{\,\gamma\max(0,\;\min(y,\,y_{\mathrm{cap}})-y_0)}\,
      H_1(y-z)
    \Big\rangle_{\mathrm{events}}, \\[2pt]
    H_1(e) =
    \begin{cases}
      e^2/2, & |e|\leq 1, \\
      |e|-1/2, & |e|>1.
    \end{cases}
  \end{gathered}
  \label{eq:loss}
\end{equation}
where $p(y)=1-10^{-y}$ is the ECDF value corresponding to the target,
and $y_{\mathrm{cap}}=5$ caps the target inside the exponent.
The factor $1+\beta p(y)^2$ gives more weight to events in the anomalous
tail. Since $p(y)$ approaches one rapidly, however, this weighting becomes
nearly constant for large $y$ and does not compensate for the decreasing
number of SM events in the deepest part of the tail. We therefore introduce
the additional factor
$10^{\gamma\max(0,\min(y,y_{\mathrm{cap}})-y_0)}$. We set $y_0=3$,
corresponding to an empirical SM tail probability of $10^{-3}$, and
$\gamma=1$. The additional weighting is therefore inactive for $y\leq3$ and
increases by a factor of ten per unit of $y$ beyond this point,
compensating for the corresponding decrease in the number of SM training
events.
The cap at $y_{\mathrm{cap}}=5$ bounds the per-event weight at about
200 if $\beta=1$. For the same reason, the loss
is accumulated in double precision: summing float32 contributions across
a batch spanning this weight range was measurably noisy, whereas float64
accumulation removed the effect.

During the adaptive-pruning stages, the total loss additionally contains the
SymbolNet sparsity regularisers for the input, coefficient, and unary-operator
gates. The value of $\beta$ is varied across the training stages as described
in Table~\ref{tab:hyper}, while $y_0=3$ and $\gamma=1$ are kept fixed.
Training uses the Adam~\cite{Kingma:2014adam} optimiser with a gradient clip norm of 1.0 and only SM
events, with the teacher scores serving as regression targets.

\section[Training and pruning procedure]{Training and pruning procedure}
\label{sec:training}

The symbolic student is trained in several stages, starting from the full
network and gradually reducing it to the final expression. The
hyperparameters used at each stage are listed in Table~\ref{tab:hyper}.

\begin{enumerate}
\item \emph{Plain training} ($\beta=1$): we first train the full
SymbolNet with all pruning masks open. This gives the model time to learn
the teacher target before any sparsity is introduced.

\item \emph{Light pruning} ($\beta=1$): we activate the trainable pruning
thresholds and impose moderate sparsity targets on the input features,
affine coefficients, and unary operators.

\item \emph{Stronger pruning} ($\beta=1$): the sparsity targets are
increased and training continues, progressively simplifying the resulting symbolic expression. The
binary product slots are kept active throughout.

\item \emph{Reduction to the target size} ($\beta=2$): the remaining
expression is reduced through a sequence of small saliency-based cuts,
each followed by a short recovery fit. The size target is set by a
teacher--student agreement criterion computed on SM events only: we
trained students at six size budgets and measured, at fixed background rates
of $10^{-3}$, $10^{-4}$, and $10^{-5}$, the fraction of teacher-selected
SM events that each student also selects
(Figure~\ref{fig:agreement-size} in Appendix~\ref{app:compression}).
The agreement at the two shallower rates saturates by roughly 500
operations, while the deepest rate continues to improve up to roughly
600 operations and flattens thereafter; 
we therefore chose
$\mathrm{CSE}\leq626$ operations, the smallest tested budget at which all three
working points have saturated. 
The evolution of the BSM performance
along the compression path (Figure~\ref{fig:compression-path} in Appendix~\ref{app:compression}) was
examined only after this choice, as a diagnostic.

The size-budget study uses SM validation and test samples to obtain
sufficient event counts at background rates down to $10^{-5}$; an
independent cross-check on an additional 7.95 million previously unused SM
events confirms the stability of the chosen budget.

\item \emph{Fixed-structure retuning} ($\beta=3$): after reaching the
target size, the pruning masks are frozen and only the surviving weights
and biases are trained. We retain the checkpoint with the lowest
regression loss on the SM validation sample.

\item \emph{Low-learning-rate retuning} ($\beta=4$): finally, we perform
one more fixed-structure fit at a lower learning rate to refine the
remaining coefficients.

\end{enumerate}

To avoid pruning parameters that are small in magnitude but might still be important for the model output, we use first-order saliency information during pruning. For the input features, we estimate the importance of $x_i$ using
\begin{equation}
\Big\langle
\left|
\frac{\partial\mathcal{L}}{\partial x_i}\,x_i
\right|
{\Big\rangle_{\mathrm{events}}},
\end{equation}
averaged over a batch of 8192 SM training events,
recompute this ranking every ten epochs, and protect the twelve most salient inputs from pruning. During the final size reduction, coefficients are ranked
using
\begin{equation}
\left|
w\,\frac{\partial\mathcal{L}}{\partial w}
\right|
\end{equation}
instead of $|w|$ alone, with the loss evaluated on one such batch.

Each epoch is trained on a fresh random subset of 300\,000 events,
independently redrawn from the 2.8M-event SM training sample at
every epoch, so that no event is systematically excluded. Because the
sparsity regularisers act on every mini-batch, the pruning schedule of
Table~\ref{tab:hyper} is calibrated to the number of gradient updates
rather than to the nominal epoch count: training on full-sample epochs at
the same epoch counts would raise the number of updates, and hence the
pruning pressure, roughly ninefold. We verified that the epoch size
affects only training throughput and not the final model, by repeating
training on the full sample with the epoch count rescaled to match the
number of gradient updates.

\begin{figure*}[!t]
  \centering
  \includegraphics[width=0.88\textwidth]{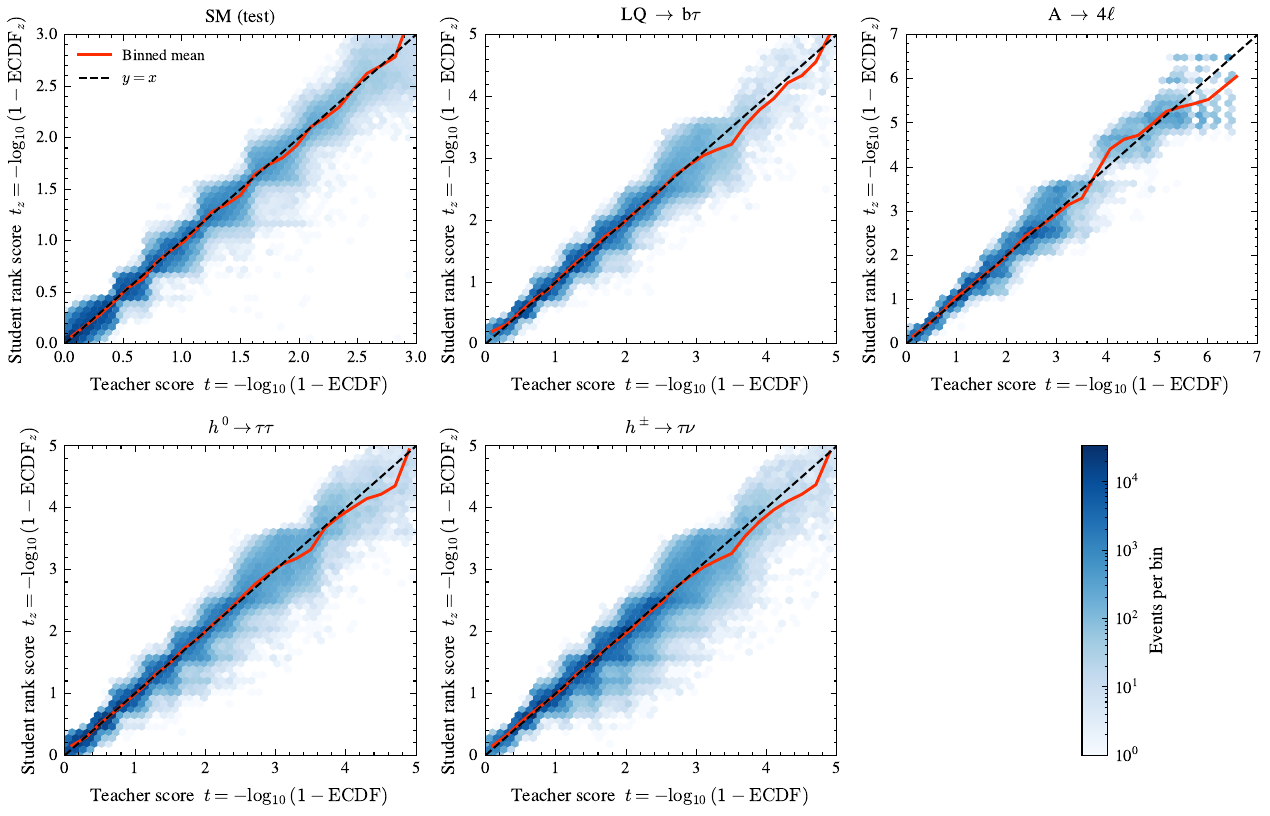}
  \caption{Event-level tail-score calibration for the symbolic student on
  the SM test sample and four BSM benchmarks. Both
  scores are mapped through their own SM-training distributions to the
  log-survival scale, with the vertical axis showing the student tail
  score $t_z=-\log_{10}(1-\mathrm{ECDF}_z)$. The colour gives the number of events per bin, the
  red curve is the binned mean for bins containing at least 20 events,
  and the dashed line denotes identical ranking. Events beyond the
  SM-training range accumulate at $\log_{10}(N+1)\simeq6.45$.}
  \label{fig:calib-student}
\end{figure*}


\begin{figure*}[!t]
  \centering
  \includegraphics[width=\textwidth]{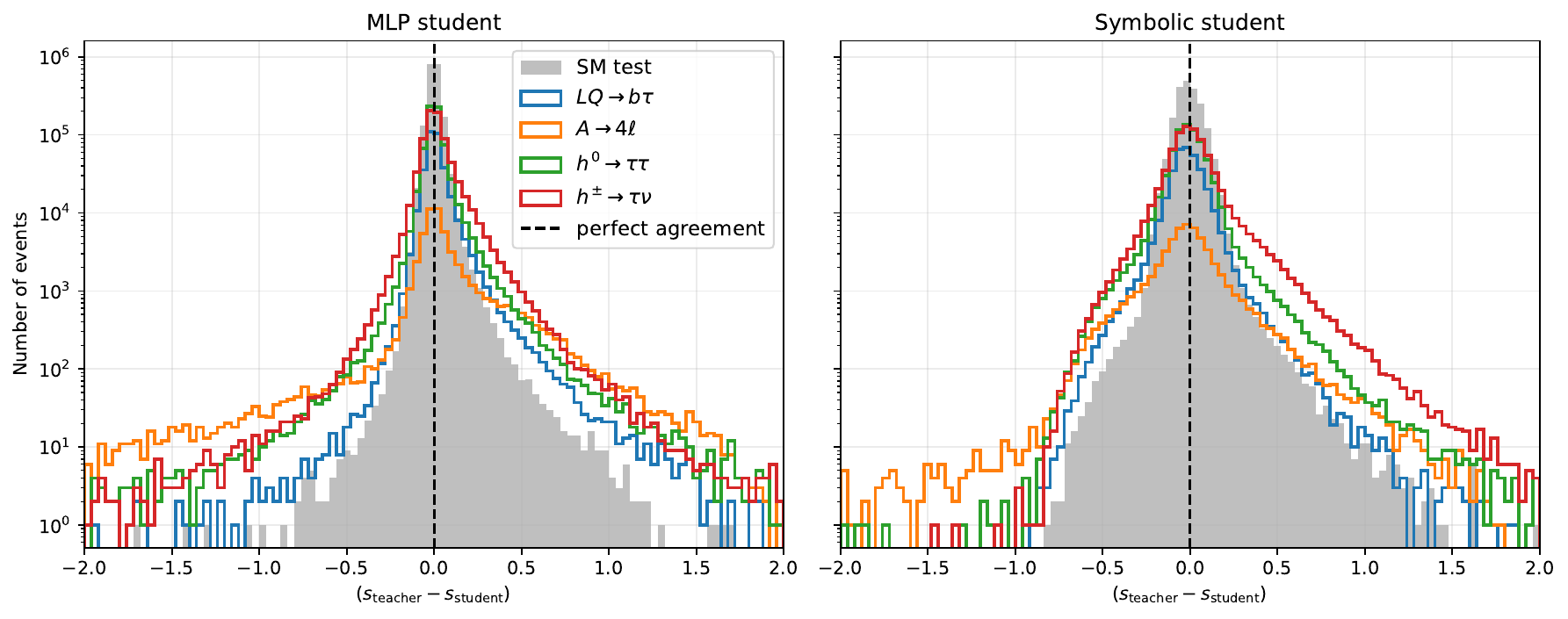}
  \caption{{Distributions of the residual
  $s_{\mathrm{teacher}} - s_{\mathrm{student}}$ for SM test events and the
  four anomaly benchmarks, for the MLP student of Ref.~\cite{companion} (left) and the symbolic student (right), where $s_{\mathrm{teacher}}$ is the teacher tail score $t$ and
  $s_{\mathrm{student}}$ the student output.}
  The dashed vertical line marks perfect agreement.}
  \label{fig:residuals-twopanel}
\end{figure*}

\section[Performance comparison]{Performance comparison}
\label{sec:performance}

\begin{figure*}[!t]
\centering
\includegraphics[width=0.62\textwidth]{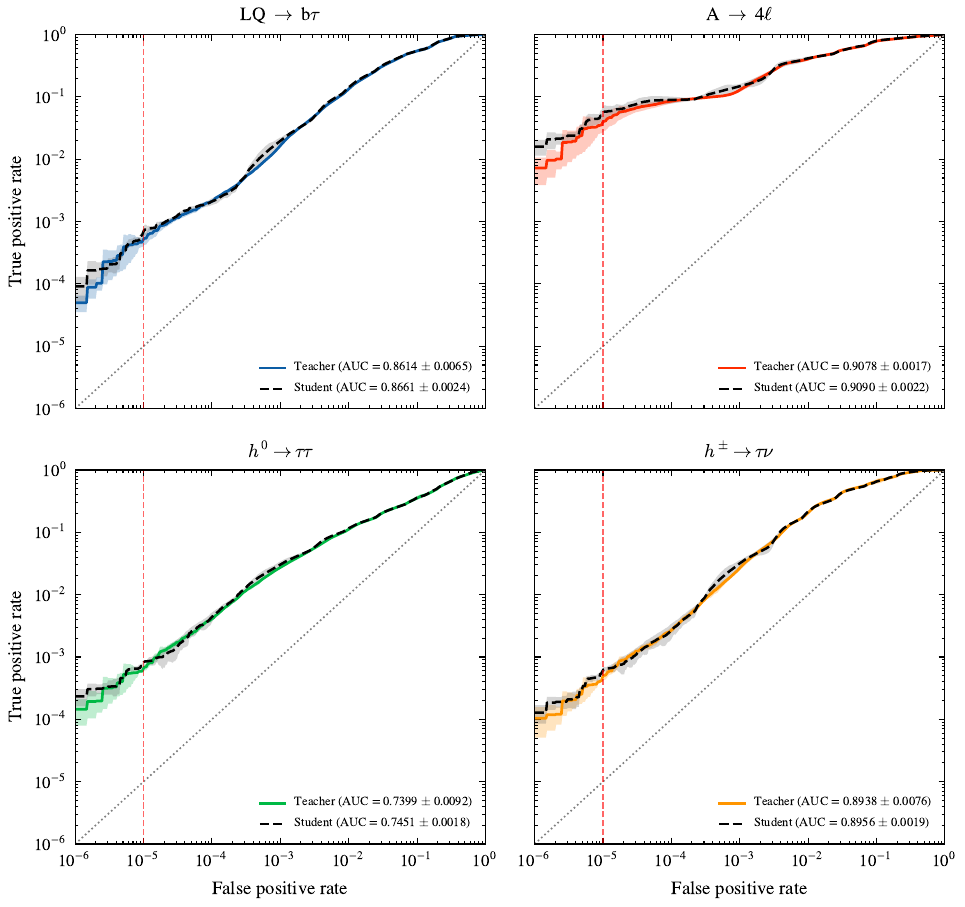}
\caption{ROC curves of the RealNVP teacher and the deployed symbolic
student on logarithmic axes for the four BSM benchmarks. The shaded
bands show the RMS deviations of four additional trainings from the
nominal curves. The
quoted AUC uncertainties are defined in the same way. The central values and uncertainties of the teacher are taken from Ref.~\cite{companion}.}
\label{fig:roc-loglog-student}
\end{figure*}

\begin{table*}[t]
\centering
\footnotesize
\setlength{\tabcolsep}{4pt}
\caption{Floating-point anomaly-detection performance of the RealNVP
teacher of Ref.~\cite{companion}, of the distilled MLP student of Ref.~\cite{companion}, and of the
symbolic student obtained in this paper on the four BSM benchmarks.  
Values correspond to the nominal training and are given in percent; uncertainties represent the RMS deviations of the
results from four additional independent trainings, performed with different random seeds, relative to
the nominal result. The value and uncertainties of the teacher and MLP student
are taken from
Ref.~\cite{companion}.}
\label{tab:results}
\begin{tabular}{llccccc}
\toprule
       &       &     & \multicolumn{4}{c}{TPR at fixed FPR} \\
\cmidrule(lr){4-7}
Signal & Model & AUC & $10^{-2}$ & $10^{-3}$ & $10^{-4}$ & $10^{-5}$ \\
\midrule
\lqbt & RealNVP teacher & 86.14${}\pm0.65$ & 13.52${}\pm0.15$ & 1.66${}\pm0.05$ & 0.21${}\pm0.01$ & 0.054${}\pm0.012$ \\
\lqbt & Distilled MLP student & 86.09${}\pm0.61$ & 13.53${}\pm0.31$ & 1.57${}\pm0.08$ & 0.19${}\pm0.02$ & 0.037${}\pm0.005$ \\
\lqbt & Symbolic student & 86.61${}\pm0.24$ & 14.26${}\pm0.30$ & 1.99${}\pm0.24$ & 0.21${}\pm0.03$ & 0.073${}\pm0.012$ \\
\midrule
\afourl & RealNVP teacher & 90.78${}\pm0.17$ & 40.52${}\pm0.91$ & 12.91${}\pm0.85$ & 8.54${}\pm0.69$ & 4.067${}\pm1.253$ \\
\afourl & Distilled MLP student & 90.66${}\pm0.18$ & 37.01${}\pm2.12$ & 12.45${}\pm1.34$ & 7.52${}\pm0.46$ & 3.104${}\pm0.230$ \\
\afourl & Symbolic student & 90.90${}\pm0.22$ & 41.59${}\pm2.69$ & 14.86${}\pm1.28$ & 8.98${}\pm1.12$ & 5.691${}\pm1.431$ \\
\midrule
\htautau & RealNVP teacher & 73.99${}\pm0.92$ & 11.13${}\pm0.25$ & 2.75${}\pm0.13$ & 0.41${}\pm0.03$ & 0.068${}\pm0.008$ \\
\htautau & Distilled MLP student & 74.00${}\pm0.93$ & 11.26${}\pm0.33$ & 2.70${}\pm0.14$ & 0.41${}\pm0.02$ & 0.054${}\pm0.008$ \\
\htautau & Symbolic student & 74.51${}\pm0.18$ & 11.62${}\pm0.21$ & 3.06${}\pm0.36$ & 0.44${}\pm0.08$ & 0.083${}\pm0.006$ \\
\midrule
\htaunu & RealNVP teacher & 89.38${}\pm0.76$ & 20.66${}\pm1.27$ & 2.70${}\pm0.21$ & 0.27${}\pm0.03$ & 0.051${}\pm0.009$ \\
\htaunu & Distilled MLP student & 89.33${}\pm0.79$ & 20.98${}\pm1.54$ & 2.70${}\pm0.16$ & 0.29${}\pm0.01$ & 0.050${}\pm0.004$ \\
\htaunu & Symbolic student & 89.56${}\pm0.19$ & 21.41${}\pm0.81$ & 3.18${}\pm0.41$ & 0.27${}\pm0.08$ & 0.063${}\pm0.006$ \\
\bottomrule
\end{tabular}
\end{table*}

The performance of the symbolic student is summarised in
Table~\ref{tab:results}, together with that of the RealNVP teacher
and of the distilled MLP student of Ref.~\cite{companion}. The
teacher is evaluated with its tail score $t$ of Eq.~(\ref{eq:ecdf}) and
the student with its output $z$. 
In order to estimate the uncertainty associated with the choice of the training starting
conditions, we vary the seed with respect to the
nominal value. 
The quoted uncertainties are the root-mean-square (RMS) deviations of
the results from four additional trainings relative to the nominal result,
as in Ref.~\cite{companion}. The symbolic student achieves discrimination
performance comparable to that of the RealNVP teacher across all four
BSM benchmarks, both in terms of AUC and in terms of true-positive rates
at the four fixed false-positive rates down to $10^{-5}$.
The event-level agreement is shown in
Figures~\ref{fig:calib-student} and~\ref{fig:residuals-twopanel}.
Figure~3 compares how the two models rank events. The student output
\(z\) is converted to a tail score \(t_z\) using the distribution of
\(z\) on SM events. The mean student tail score \(t_z\) follows the
teacher tail score \(t\) across the full range in all five samples.
Figure~4 compares the scores directly. The residuals between \(z\)
and \(t\) are narrow and centred at zero, with a larger spread in the
sparsely populated high-score tail. 
The MLP student of Ref.~\cite{companion}, shown with
the symbolic student in Figure~\ref{fig:residuals-twopanel}, has a narrower residual core, whereas the
symbolic student has fewer large residuals on \afourl{} events.

The comparison with the distilled MLP of Ref.~\cite{companion} is
particularly direct, since both students use the same frozen teacher, the
same distillation target, and the same data. 
Across the four benchmarks, the symbolic student achieves AUCs
comparable to those of the distilled MLP and broadly comparable
signal efficiencies at fixed false-positive rates.


The low-background region is shown in
Figure~\ref{fig:roc-loglog-student}, where the ROC curves are plotted on
logarithmic axes, with the variation across retrainings shown as a
band. The symbolic student follows the teacher closely over the
full range of false-positive rates considered.

The compression is also visible at the level of the input representation.
Although all 57 trigger-level features are available during training, only
30 appear in the final symbolic expression. Since the student is an explicit
formula, the retained inputs can be identified directly rather than inferred
from an attribution method. Their distribution across the reconstructed
physics objects and kinematic variables is shown in
Figure~\ref{fig:feature-usage} in Appendix~\ref{app:features}.

\section{FPGA implementation and resource comparison}
\label{sec:fpga}

The deployed expression of Appendix~\ref{app:formula} is turned into
firmware with the symbolic-expression backend of
hls4ml~\cite{Duarte:2018ite,Schulte:2025mai,Tsoi:2023isc} (version 1.3.0),
which writes the expression directly into the high-level-synthesis (HLS)
C++ source, with the coefficients as constants of the arithmetic. All
inputs, coefficients, intermediate values and the output share one
fixed-point type, \texttt{ap\_fixed<20,10>}: 20 bits, of which 10 are
integer bits including the sign, with convergent rounding and saturation
on overflow. This is the narrowest type for which, in a scan on the
bit-accurate emulation of the generated C++, the AUC and the true-positive
rates at background rates of $10^{-3}$ to $10^{-5}$ stay within
0.1 percentage points and 10\%, respectively, of their floating-point
values on every benchmark; with fewer integer bits the function arguments
of the most anomalous events saturate and the deep-tail efficiency drops
while the AUC is unchanged.

The hyperbolic tangent, the Gaussian $e^{-u^2}$ and the cosine are read
from 512-entry lookup tables, and the three remaining exponentials are
evaluated with the HLS library.
Each of the 280 multiplicative coefficients is rounded after training to
the nearest value with at most two non-zero signed binary digits, so that
every multiplication by a constant reduces to at most one shift-and-add;
this lowers the number of adders for the constant multiplications from
600 to 255. Table~\ref{tab:fixedpoint} compares the floating-point
expression with the bit-accurate emulation of the resulting fixed-point
C++ on the full test statistics: the AUC changes by at most 0.13
percentage points, and the true-positive rates at background rates of
$10^{-3}$ to $10^{-5}$ stay within 5\% of their floating-point values.

\begin{table*}[t]
\centering
\small
\setlength{\tabcolsep}{6pt}
\caption{Anomaly-detection performance of the deployed symbolic student in
floating-point arithmetic (as in Table~\ref{tab:results}) and of its
fixed-point firmware (\texttt{ap\_fixed<20,10>}, coefficients rounded to
two signed digits), evaluated with the bit-accurate emulation of the
generated C++ on the full test statistics. All values are given in
percent.}
\label{tab:fixedpoint}
\begin{tabular}{llccccc}
\toprule
       &       &     & \multicolumn{4}{c}{TPR at fixed FPR} \\
\cmidrule(lr){4-7}
Signal & Arithmetic & AUC & $10^{-2}$ & $10^{-3}$ & $10^{-4}$ & $10^{-5}$ \\
\midrule
\lqbt & Floating point & 86.61 & 14.26 & 1.99 & 0.210 & 0.073 \\
\lqbt & Fixed-point firmware & 86.48 & 14.22 & 2.03 & 0.210 & 0.070 \\
\midrule
\afourl & Floating point & 90.90 & 41.59 & 14.86 & 8.976 & 5.691 \\
\afourl & Fixed-point firmware & 90.89 & 41.67 & 15.16 & 8.998 & 5.578 \\
\midrule
\htautau & Floating point & 74.51 & 11.62 & 3.06 & 0.439 & 0.083 \\
\htautau & Fixed-point firmware & 74.46 & 11.52 & 3.02 & 0.439 & 0.080 \\
\midrule
\htaunu & Floating point & 89.56 & 21.41 & 3.18 & 0.268 & 0.063 \\
\htaunu & Fixed-point firmware & 89.46 & 21.15 & 3.15 & 0.267 & 0.060 \\
\bottomrule
\end{tabular}
\end{table*}

The firmware is synthesised with AMD Vitis HLS 2025.1 as a fully
pipelined design that accepts one event per clock cycle, and is taken
through out-of-context (OOC) synthesis, placement and routing with
Vivado 2025.1 targeting the Xilinx Virtex UltraScale+ VU13P
(\texttt{xcvu13p-flga2577-2-e}) at a 200\,MHz clock, the device and clock
target of Ref.~\cite{companion}, with resource and timing reported from the tool report. The routed design meets the 200\,MHz timing target. Its estimated post-route maximum clock frequency is 210.5\,MHz, corresponding to a clock period of 4.751\,ns. The nine-cycle latency is 42.8\,ns at this frequency and 45\,ns at the 200\,MHz target frequency. The routed design uses 14\,049 lookup tables (LUTs), 4\,785 flip-flops (FFs), 54 DSP slices, and 13 block RAMs, below 1\% of the device in each category.

Table~\ref{tab:resources} places this result next to the MLP student of
Ref.~\cite{companion}, implemented on the same device and with the same
clock target as multiplier-free Verilog at three bit widths. The symbolic
student needs 3.9 times fewer LUTs and 5.0 times fewer flip-flops than
the 8-bit MLP, the configuration that preserves the floating-point
performance of that student, and fewer LUTs than even its 3-bit
configuration.
Unlike the MLP designs, it uses DSP slices, for the products between two
variables and inside the exponential and cosine evaluations, and block
RAMs, for the function tables.
Since the two students are also implemented with different flows,
multiplier-free Verilog for the MLP and HLS for the symbolic student, the
comparison indicates a favourable trade-off between resources and
performance rather than an advantage in every hardware resource.

\begin{table*}[t]
\centering
\caption{Post-route timing and resource usage on the Xilinx Virtex
UltraScale+ VU13P. Latencies are reported at each design's post-route
$F_{\max}$, following Ref.~\cite{companion}. At the 200\,MHz target
clock frequency, the symbolic student's nine-cycle latency is 45\,ns.}
\label{tab:resources}
\begin{tabular}{llrrrrrr}
\toprule
Model & Precision & $F_{\max}$ [MHz] & Latency at $F_{\max}$ [ns]
      & DSP & LUT & FF & BRAM \\
\midrule
MLP student~\cite{companion} & 3-bit & 221.1 & 27.1 & 0 & 15\,720 & 8\,160 & 0 \\
MLP student~\cite{companion} & 4-bit & 222.2 & 27.0 & 0 & 20\,954 & 6\,867 & 0 \\
MLP student~\cite{companion} & 8-bit & 211.8 & 51.9 & 0 & 54\,740 & 23\,693 & 0 \\
\midrule
Symbolic student (this work) & 20-bit, 2-digit coefficients
                             & 210.5 & 42.8 & 54 & 14\,049 & 4\,785 & 13 \\
\bottomrule
\end{tabular}
\end{table*}

\section{Conclusions}
\label{sec:summary}

We have presented a pipeline for unsupervised anomaly detection at the LHC
Level-1 Trigger in which the anomaly scores of the RealNVP
normalising-flow teacher of Ref.~\cite{companion} are distilled into a
closed-form symbolic expression. Using SymbolNet~\cite{Tsoi:2024ypg}, the
student learns both the functional form of the expression and a sparse
subset of the available trigger-level inputs. The final symbolic student
contains 577 operations after common-subexpression elimination and depends
on 30 of the 57 input features.

The symbolic student retains discrimination performance comparable to that of the substantially larger teacher across all four BSM benchmarks, including at false-positive rates as low as $10^{-5}$.
All training,
pruning, checkpoint selection, and other model-development choices are made
using SM events only.

Because the student is an explicit formula rather than a network, its
arithmetic complexity is fixed and countable before any synthesis step.
In 20-bit fixed-point arithmetic with every coefficient rounded to two
signed binary digits, the expression retains its floating-point
performance to within 0.13 percentage points in AUC. 
The routed firmware on a Xilinx Virtex UltraScale+ VU13P processes one event per clock cycle and has a nine-cycle latency, corresponding to 42.8\,ns at the estimated post-route maximum clock frequency. It uses 14\,049 lookup tables, 4\,785 flip-flops, 54 DSP slices and 13 block RAMs.

The closest comparison is with the MLP student of Ref.~\cite{companion},
which is distilled from the same frozen teacher using the same target and
data partition. The symbolic student achieves performance comparable to that of the
teacher and the distilled MLP across the four benchmarks. The symbolic student
reads 30 rather than 57 inputs and contains 577 operations against the
7\,808 dense weight multiplications of the MLP. On the same device
and at the same clock target, its firmware uses 3.9 times fewer
lookup tables and 5.0 times fewer flip-flops than the 8-bit MLP
student, at a latency within the range of the MLP designs and at the cost
of a small number of DSP slices and block RAMs, a favourable
trade-off between resources and performance.

More broadly, these results show that knowledge distillation can separate
the complexity of the model used to learn the SM distribution from the
representation ultimately used for real-time inference. Symbolic regression
takes this separation one step further: rather than deploying another neural
network, the resulting trigger algorithm is an explicit mathematical
expression whose inputs, operations, and numerical behaviour can be examined
directly. The framework is not tied to the particular RealNVP teacher used
here; in principle, another model producing a scalar anomaly score could be
substituted and a new symbolic student trained within the same distillation
and deployment pipeline. This provides a route for translating increasingly
expressive anomaly-detection models into compact and transparent
representations compatible with the constraints of real-time triggering.

\section*{Funding}

Roope Oskari Niemi acknowledges funding from the Eric \& Wendy Schmidt Fund
for Strategic Innovation through the CERN Next Generation Triggers project
under Grant Agreement No.~SIF-2023-004.\\

\noindent{Chang Sun is partially supported by United States DoE (grant numbers
DE-SC0011925, DE-FOA-0002705) and NSF (grant numbers PHY240298, PHY2117997).}

\section*{Acknowledgements}
A first draft of this paper was generated with the assistance of Anthropic’s Claude, based on existing code, a technical document, and a set of publications on related FPGA toolflows. The draft was subsequently reviewed and edited by the authors. OpenAI ChatGPT has been used for a final editorial check.
Anthropic's Claude and OpenAI ChatGPT were also used, under the authors’ supervision, to assist with code development. All AI-assisted code was reviewed and validated by the authors.

\section*{Data availability}
The LHC benchmark datasets used in this work are publicly available on
Zenodo~\cite{zenodo:lq,zenodo:a4l,zenodo:h0tt,zenodo:hctaunu}. The deployed
expression is given in Appendix~\ref{app:formula} and, at full
precision, at \url{https://github.com/jusbol/symbolic-student-expression}.

\bibliographystyle{auto_generated}
\bibliography{bibliography}

\clearpage
\appendix

\onecolumn
\section{Choice of the expression size and compression path}
\label{app:compression}

\begin{figure}[!ht]
\centering
\includegraphics[width=0.55\textwidth]{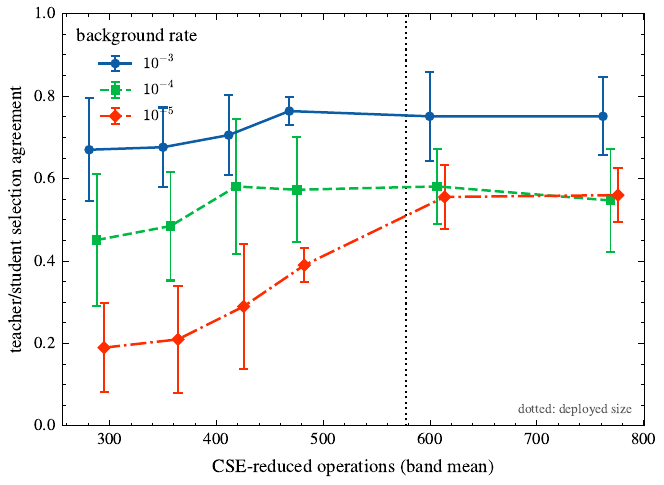}
\caption{Teacher--student selection agreement versus expression size, on
SM events only. Complete students were trained at six size budgets (five
retrainings per budget, nine at the chosen one); each point is placed at
the mean operation count of its budget, and the error bars give the
spread across retrainings. At each background rate, teacher and student
each select the same fraction of SM test events, and the vertical axis
shows the fraction of the teacher-selected events that the student also
selects. The dotted line marks the deployed size. The size target
$\mathrm{CSE}\leq626$ is the smallest budget at which all three curves
have flattened, with the deepest working point, $10^{-5}$, saturating
last.}
\label{fig:agreement-size}
\end{figure}

\begin{figure}[!ht]
\centering
\includegraphics[width=0.55\textwidth]{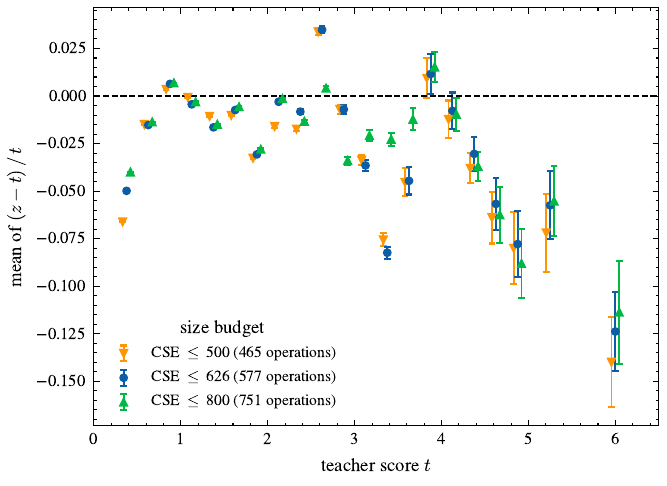}
\caption{Mean relative residual $(z-t)/t$ between the student
output $z$ and the teacher score $t$ as a function of $t$, on SM test
events, for the selected size budget ($\mathrm{CSE}\leq626$) and for the
neighbouring smaller and larger budgets, obtained from the same training.
The error bars give the statistical uncertainty of the mean in each bin.
The region $t<0.25$, where the denominator approaches zero, is not
shown.}
\label{fig:profile-residual}
\end{figure}

\begin{figure}[!ht]
\centering
\includegraphics[width=0.55\textwidth]{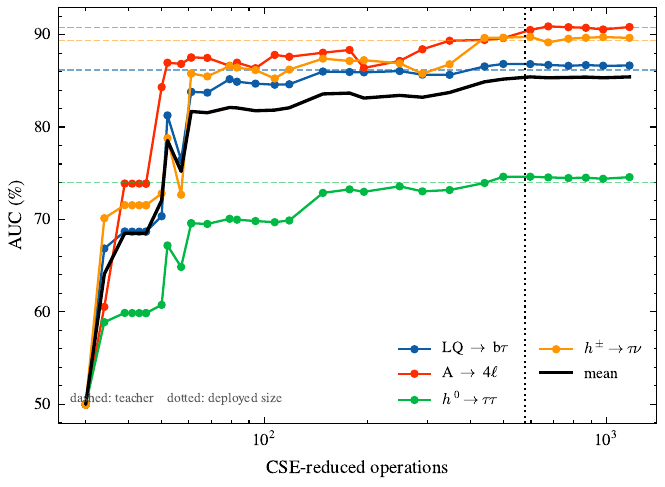}
\caption{AUC versus expression size along the symbolic student's
compression path for the four BSM benchmarks and their mean. Each point
is evaluated after a pruning step and its short recovery fit. The star
marks the fully retuned deployed expression, while the dashed horizontal
line shows the teacher performance. Sizes are CSE-reduced operation counts.}
\label{fig:compression-path}
\end{figure}

The expression size was fixed before any signal sample was examined.
Figure~\ref{fig:agreement-size} shows the criterion: teacher--student
selection agreement on SM events as a function of the size budget. The
agreement at background rates of $10^{-3}$ and $10^{-4}$ saturates by
roughly 500 operations, while the $10^{-5}$ working point continues to
improve up to roughly 600 operations; the target $\mathrm{CSE}\leq626$
is the smallest budget beyond which no working point improves further.

The error bars in Figure~\ref{fig:agreement-size} give the spread
across retrainings: at the chosen budget the agreement is $0.75\pm0.11$
at a background rate of $10^{-3}$, $0.58\pm0.09$ at $10^{-4}$, and
$0.56\pm0.08$ at $10^{-5}$.

As a further cross-check, Figure~\ref{fig:profile-residual} shows the
mean relative residual between the student output and the teacher score
as a function of the teacher score, for the chosen budget and its two
neighbours. In the high-score region, $t\gtrsim4$, the chosen and the
larger budget agree within the statistical uncertainties, while the
smaller budget lies slightly lower in the two highest bins.

Figure~\ref{fig:compression-path} traces the compression path followed during
the size-targeted pruning stages. The AUC values shown here were obtained
only \emph{after} the size target had been fixed with the SM-only criterion
above; the figure is reported as a diagnostic of the pruning procedure, and
no BSM information entered the size choice. The deployed model is the fully
retuned expression marked by the star.

\section{Input features of the deployed expression}
\label{app:features}

\begin{figure}[!ht]
\centering
\includegraphics[width=0.55\textwidth]{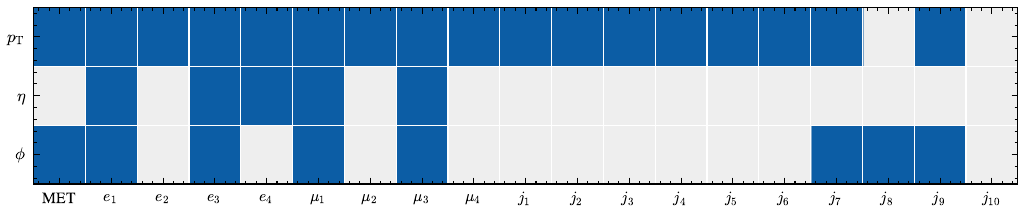}
\caption{Input features appearing in the deployed symbolic expression, in the
standard dataset ordering: 19 reconstructed objects (missing transverse
energy, four electrons, four muons, ten jets), each described by
$(p_{\mathrm{T}}, \eta, \phi)$. Filled cells are read by the expression;
30 of the 57 available inputs survive. The $\eta$ component of the missing
transverse energy is identically zero in the dataset and is never selected.}
\label{fig:feature-usage}
\end{figure}

Figure~\ref{fig:feature-usage} shows which inputs the deployed expression
reads. The selection concentrates on transverse momenta and retains the
missing-transverse-energy magnitude, consistent with the saliency ranking
observed during training.

\section{Closed-form symbolic-student expression}
\label{app:formula}

The complete deployed formula of the symbolic student is reproduced
below with coefficients rounded to three significant digits ($x_i$ denotes
the $i$-th standardised input feature; the deployed version uses full
precision). The complete expression, with full-precision coefficients
and the input standardisation constants, is available at \url{https://github.com/jusbol/symbolic-student-expression}. Common subexpressions are named $u_j$; the final sum is split
into the print-only groups $z_{\mathrm{A}}$ and $z_{\mathrm{B}}$; and
$\operatorname{gauss}(u)$ denotes $e^{-u^2}$; every square is evaluated
with its argument clipped to $[-10,10]$, as in training
(Section~\ref{sec:student}). The displayed equations are the entire
trigger algorithm:
\input{_appendix_expr.tex}
The output $z$ approximates the score $t = -\log_{10}(1-\mathrm{ECDF})$
and is thresholded directly. The deployed full-precision form has 577
operations after common-subexpression elimination.

\end{document}

%% file: _appendix_expr.tex
\begingroup
\setlength{\columnsep}{1.6em}
\setlength{\multicolsep}{4pt}
\setlength{\parindent}{0pt}
\setlength{\parskip}{0pt}
\setlength{\emergencystretch}{3em}
\footnotesize
\binoppenalty=0
\relpenalty=0
\hbadness=10000
\def\Eline#1{\par\hangindent=1.2em\hangafter=1\noindent$#1$}
\begin{multicols}{2}
\raggedcolumns
\raggedright
\Eline{u_{0} = \operatorname{gauss}\bigl(0.646 - 1.15 x_{33}\bigr)}
\Eline{u_{1} = \tanh\bigl(- 0.124 x_{21} - 0.269 x_{3} + 0.184 x_{30} + 0.145 x_{36} + 0.269 x_{6}\bigr)}
\Eline{u_{2} = \tanh\bigl(0.348 x_{36} + 0.378 x_{39} + 0.0932 x_{42} + 1.7 x_{6} - 0.375 x_{9} + 0.207\bigr)}
\Eline{u_{3} = \tanh\bigl(0.504 x_{18} - 0.000695 x_{27} - 0.175 x_{3} + 0.121 x_{30} - 0.154 x_{33} + 1.35 x_{6} + 0.16\bigr)}
\Eline{u_{4} = \cos\bigl(- 0.233 x_{10} + 0.383 x_{13} + 0.479 x_{22} - 0.189 x_{23} - 0.203 x_{24} + 0.455 x_{30} + 0.15 x_{51}\bigr)}
\Eline{u_{5} = \tanh\bigl(0.0363 x_{15} + 0.0452 x_{3} + 0.0483 x_{30} + 0.647 x_{33} + 0.538 x_{36} + 0.763 x_{39} - 0.162 x_{9} - 0.389\bigr)}
\Eline{u_{6} = (0.235 x_{17} + 0.249 x_{5})^{2}}
\Eline{u_{7} = \cos\bigl(- 0.294 x_{22} - 0.281 x_{24} - 0.334 x_{4} - 0.155 x_{45} - 0.147 x_{47} + 0.304 x_{50} - 0.194 x_{51} - 0.124 x_{53} + 0.164 x_{9}\bigr)}
\Eline{u_{8} = x_{36}^{2}}
\Eline{u_{9} = \cos\bigl(- 0.147 x_{10} + 0.117 x_{11} + 0.138 x_{15} + 0.218 x_{21} + 0.3 x_{22} - 0.263 x_{3} - 0.31 x_{39} + 0.153 x_{50} + 0.00512\bigr)}
\Eline{u_{10} = \cos\bigl(- 0.426 x_{0} - 0.424 x_{10} + 0.0999 x_{2} + 0.136 x_{21}\bigr)}
\Eline{u_{11} = \operatorname{gauss}\bigl(1.18 x_{6} - 0.738\bigr)}
\Eline{u_{12} = \operatorname{gauss}\bigl(- 0.00296 x_{16} + 1.21 x_{3} - 0.88 x_{6} + 0.415\bigr)}
\Eline{u_{13} = (0.693 x_{15} + 0.45 x_{18}) (- 0.584 x_{3} - 0.481 x_{6} - 0.22)}
\Eline{u_{14} = (0.543 x_{15} + 0.592 x_{18} + 0.028 x_{36}) (- 0.0095 x_{0} - 0.573 x_{3} - 0.0969 x_{39} - 0.763 x_{6})}
\Eline{u_{15} = \operatorname{gauss}\bigl(1.26 x_{15} + 0.425\bigr)}
\Eline{u_{16} = \cos\bigl(0.31 x_{10} - 0.262 x_{11} + 0.141 x_{21} - 0.471 x_{22}\bigr)}
\Eline{u_{17} = \tanh\bigl(0.0542 x_{15} + 1.79 x_{18} - 0.227 x_{3} + 0.341 x_{39} + 0.0895\bigr)}
\Eline{u_{18} = \operatorname{gauss}\bigl(- 1.48 x_{30}\bigr)}
\end{multicols}
\medskip
\Eline{z = z_{A} + z_{B}}
\Eline{z_{A} = 0.0271 u_{2} + 0.0443 u_{3} + 0.0532 u_{5} + 0.00287 x_{12} + 0.00352 x_{18} + 0.000643 x_{21} + 0.000944 x_{3} + 0.00237 x_{51} + 0.00441 x_{6} + 0.137 \operatorname{gauss}\bigl(- 0.676 u_{6} - 0.067 x_{15} - 1.67 x_{27} - 0.0902 x_{3} - 0.156 x_{30} - 0.777\bigr) + 0.0936 \operatorname{gauss}\bigl(- 0.582 u_{10} + 0.159 u_{12} - 0.312 u_{13} - 0.152 u_{14} + 0.157 u_{4} + 0.477 u_{6} + 0.0736 u_{7} + 0.213 x_{15} + 0.0653 x_{2} x_{5} + 0.368 x_{27} + 0.287 x_{3} + 0.495 x_{30} + 0.278\bigr) + 0.132 \operatorname{gauss}\bigl(- 0.495 u_{1} - 0.182 x_{12} + 0.00742 x_{15} - 0.223 x_{18} - 0.0408 x_{21} - 0.235 x_{24} - 0.0599 x_{3} - 0.0561 x_{42} - 0.104 x_{45} + 0.000993 x_{50}^{2} - 0.15 x_{51} - 0.28 x_{6} - 0.0606 x_{9} - 0.043\bigr) + 0.617 \tanh\bigl(0.0315 u_{0} + 0.0158 u_{12} - 0.356 u_{13} - 0.417 u_{14} + 0.14 u_{15} + 0.0602 u_{16} + 0.103 u_{18} + 0.156 u_{4} + 0.151 u_{5} + 0.0974 u_{7} + 0.126 x_{12} + 0.281 x_{15} + 0.235 x_{18} + 0.0283 x_{21} + 0.163 x_{24} + 0.0787 x_{27} + 0.267 x_{3} + 0.217 x_{30} + 0.0302 x_{33} + 0.0389 x_{42} + 0.0719 x_{45} + 0.104 x_{51} + 0.316 x_{6} + 0.201 x_{9} + 0.216\bigr) + 0.601 \tanh\bigl(0.0535 u_{11} - 0.0808 u_{13} - 0.177 u_{14} + 0.0814 u_{15} - 0.158 u_{16} + 0.351 u_{17} + 0.0655 u_{18} - 0.176 u_{1} + 0.356 u_{2} + 0.0793 u_{4} - 0.154 u_{7} - 0.00114 u_{8} + 0.102 x_{12} + 0.108 x_{15} - 0.0573 x_{18} + 0.0228 x_{21} + 0.131 x_{24} + 0.00613 x_{27} + 0.127 x_{3} + 0.0533 x_{30} + 0.0413 x_{33} + 0.0342 x_{36} + 0.0399 x_{4} (0.433 x_{16} + 0.306 x_{4}) + 0.0313 x_{42} + 0.0579 x_{45} + 0.0838 x_{51} + 0.0431 x_{6} + 0.117 x_{9} + 0.262 \tanh\bigl(- 0.00304 x_{30} + 0.913 x_{36} + 0.13 x_{9}\bigr) - 0.0852\bigr) + 0.329}
\Eline{z_{B} = 0.0153 u_{0} - 0.0454 u_{1} - 0.00717 u_{4} - 0.00972 u_{7} - 0.000117 x_{15} + 0.0037 x_{24} + 0.000884 x_{42} + 0.00163 x_{45} + 0.000955 x_{9} - 0.0569 \operatorname{gauss}\bigl(0.636 u_{1} - 0.332 u_{4} + 0.00214 u_{8} - 0.284 u_{9} - 0.166\bigr) - 0.732 \tanh\bigl(0.159 u_{10} - 0.126 u_{3} - 0.122 u_{6} - 0.0183 x_{12} + 0.000748 x_{15} - 0.0225 x_{18} - 0.00412 x_{21} - 0.0237 x_{24} - 0.00604 x_{3} - 0.00566 x_{42} - 0.0105 x_{45} - 0.0151 x_{51} - 0.0282 x_{6} - 0.00612 x_{9} - 0.00434\bigr) + 0.457 \tanh\bigl(0.385 u_{11} + 0.194 u_{1} + 0.131 u_{4} + 0.245 x_{12} - 0.00999 x_{15} + 0.301 x_{18} + 0.055 x_{21} + 0.316 x_{24} + 0.0807 x_{3} + 0.0756 x_{42} + 0.14 x_{45} + 0.202 x_{51} + 0.377 x_{6} + 0.0817 x_{9} + 0.0579\bigr) - 0.748 \tanh\bigl(- 0.168 u_{0} + 0.0545 u_{15} + 0.186 u_{16} - 0.294 u_{17} - 0.00532 u_{1} + 0.171 u_{3} - 0.0886 u_{5} + 0.115 u_{7} - 0.066 u_{9} - 0.0187 x_{12} + 0.000765 x_{15} - 0.023 x_{18} - 0.00421 x_{21} - 0.0242 x_{24} - 0.00617 x_{3} - 0.00578 x_{42} - 0.0107 x_{45} - 0.0155 x_{51} - 0.0288 x_{6} - 0.00625 x_{9} + 0.232\bigr) - 0.568 \tanh\bigl(- 0.154 u_{0} + 0.447 u_{10} - 0.196 u_{11} - 0.612 u_{12} - 0.42 u_{15} - 0.229 u_{16} - 0.37 u_{17} - 0.308 u_{18} - 0.184 u_{4} - 0.167 u_{5} - 0.0874 u_{7} - 0.326 x_{12} - 0.18 x_{15} - 0.39 x_{18} - 0.0731 x_{21} - 0.42 x_{24} - 0.334 x_{27} - 0.355 x_{3} - 0.449 x_{30} + 0.0208 x_{33} + 0.028 x_{36} - 0.1 x_{42} - 0.186 x_{45} - 0.269 x_{51} - 0.501 x_{6} - 0.109 x_{9} - 0.429 (0.339 x_{16} - 0.207 x_{4})^{2} - 0.425\bigr) .}
\endgroup

%% file: bibliography.bib
@article{ATLAS:2008xda,
  collaboration = "ATLAS",
  title         = "{The ATLAS Experiment at the CERN Large Hadron Collider}",
  doi           = "10.1088/1748-0221/3/08/S08003",
  url           = "https://doi.org/10.1088/1748-0221/3/08/S08003",
  journal       = "JINST",
  volume        = "3",
  pages         = "S08003",
  year          = "2008"

}

@article{CMS:2008xjf,
  collaboration = "CMS",
  title         = "{The CMS Experiment at the CERN LHC}",
  doi           = "10.1088/1748-0221/3/08/S08004",
  url           = "https://doi.org/10.1088/1748-0221/3/08/S08004",
  journal       = "JINST",
  volume        = "3",
  pages         = "S08004",
  year          = "2008"
}

@article{CMS:2020cmk,
  collaboration = "CMS",
  title         = "{Performance of the CMS Level-1 trigger in proton-proton collisions at $\sqrt{s}=13$ TeV}",
  doi           = "10.1088/1748-0221/15/10/P10017",
  url           = "https://doi.org/10.1088/1748-0221/15/10/P10017",
  journal       = "JINST",
  volume        = "15",
  number        = "10",
  pages         = "P10017",
  year          = "2020"
}

@techreport{CMS:L1TDR,
  collaboration = "CMS",
  title         = "{The Phase-2 Upgrade of the CMS Level-1 Trigger}",
  reportNumber  = "CERN-LHCC-2020-004, CMS-TDR-021",
  institution   = "{CERN}",
  year          = "2020",
  note          = {\href{https://cds.cern.ch/record/2714892}{\texttt{cds:2714892}}}
}

@article{Kasieczka:2021xcg,
  author        = "Kasieczka, Gregor and others",
  title         = "{The LHC Olympics 2020: a community challenge for anomaly detection in high energy physics}",
  eprint        = "2101.08320",
  archivePrefix = "arXiv",
  primaryClass  = "hep-ph",
  doi           = "10.1088/1361-6633/ac36b9",
  url           = "https://arxiv.org/abs/2101.08320",
  journal       = "Rept. Prog. Phys.",
  volume        = "84",
  number        = "12",
  pages         = "124201",
  year          = "2021"
}

@article{Aarrestad:2021oeb,
  author        = "Aarrestad, T. and others",
  title         = "{The Dark Machines Anomaly Score Challenge}",
  eprint        = "2105.14027",
  archivePrefix = "arXiv",
  primaryClass  = "hep-ph",
  doi           = "10.21468/SciPostPhys.12.1.043",
  url           = "https://arxiv.org/abs/2105.14027",
  journal       = "SciPost Phys.",
  volume        = "12",
  number        = "1",
  pages         = "043",
  year          = "2022"
}

@article{Kingma:2013vae,
  author        = "Kingma, Diederik P. and Welling, Max",
  title         = "{Auto-Encoding Variational Bayes}",
  eprint        = "1312.6114",
  archivePrefix = "arXiv",
  primaryClass  = "stat.ML",
  year          = "2013"
}

@article{Cerri:2018anp,
  author        = "Cerri, Olmo and Nguyen, Thong Q. and Pierini, Maurizio and Spiropulu, Maria and Vlimant, Jean-Roch",
  title         = "{Variational autoencoders for new physics mining at the Large Hadron Collider}",
  eprint        = "1811.10276",
  archivePrefix = "arXiv",
  primaryClass  = "hep-ex",
  doi           = "10.1007/JHEP05(2019)036",
  url           = "https://arxiv.org/abs/1811.10276",
  journal       = "JHEP",
  volume        = "05",
  pages         = "036",
  year          = "2019"
}

@article{Knapp:2020adversarial,
  author        = "Knapp, Oliver and others",
  title         = "{Adversarially Learned Anomaly Detection on CMS Open Data: re-discovering the top quark}",
  eprint        = "2005.01598",
  archivePrefix = "arXiv",
  primaryClass  = "hep-ex",
  year          = "2020"
}

@article{Govorkova:2021utb,
  author        = "Govorkova, Ekaterina and others",
  title         = "{Autoencoders on FPGAs for real-time, unsupervised new physics detection at 40 MHz at the LHC}",
  eprint        = "2108.03986",
  archivePrefix = "arXiv",
  primaryClass  = "physics.ins-det",
  doi           = "10.1038/s42256-022-00441-3",
  url           = "https://arxiv.org/abs/2108.03986",
  journal       = "Nature Mach. Intell.",
  volume        = "4",
  number        = "2",
  pages         = "154--161",
  year          = "2022"
}

@article{Gandrakota:2024axoltl,
  author        = "Gandrakota, Adarsh",
  title         = "{Real-time Anomaly Detection at the L1 Trigger of CMS Experiment}",
  eprint        = "2411.19506",
  archivePrefix = "arXiv",
  primaryClass  = "hep-ex",
  year          = "2024"
}

@article{Hinton:2015distill,
  author        = "Hinton, Geoffrey and Vinyals, Oriol and Dean, Jeff",
  title         = "{Distilling the Knowledge in a Neural Network}",
  eprint        = "1503.02531",
  archivePrefix = "arXiv",
  primaryClass  = "stat.ML",
  year          = "2015"
}

@misc{Cohen:2025gelato,
  author        = "Cohen, M. M. and others",
  title         = "{GELATO: a Generic Event-Level Anomalous Trigger Option for ATLAS}",
  howpublished  = "CERN CDS record",
  note          = {\href{https://cds.cern.ch/record/2938881}{\texttt{cds:2938881}}},
  year          = "2025"
}

@article{Vaselli:2025fad,
    author = "Vaselli, Francesco and Sun, Chang and Aarrestad, Thea and Danopoulos, Dimitrios and Oskari Niemi, Roope and Glowacki, Maciej Mikolaj and Govorkova, Katya and Loncar, Vladimir and Pantaleo, Felice and Pierini, Maurizio",
    title = "{It{\textquoteright}s not a FAD: first demonstration of flows for unsupervised anomaly detection at 40{\,}MHz for use at the Large Hadron Collider}",
    eprint = "2508.11594",
    archivePrefix = "arXiv",
    primaryClass = "hep-ex",
    doi = "10.1088/2632-2153/ae51dd",
    journal = "Mach. Learn. Sci. Tech.",
    volume = "7",
    number = "2",
    pages = "025052",
    year = "2026"
}

@article{Kingma:2014adam,
  author        = "Kingma, Diederik P. and Ba, Jimmy",
  title         = "{Adam: A Method for Stochastic Optimization}",
  eprint        = "1412.6980",
  archivePrefix = "arXiv",
  primaryClass  = "cs.LG",
  year          = "2014"
}

@article{Govorkova:2021dataset,
  author        = "Govorkova, Ekaterina and others",
  title         = "{LHC physics dataset for unsupervised New Physics detection at 40 MHz}",
  eprint        = "2107.02157",
  archivePrefix = "arXiv",
  primaryClass  = "physics.data-an",
  year          = "2021"
}

@dataset{Aarrestad:2021v2,
  author    = "Aarrestad, Thea and Govorkova, Ekaterina and Ngadiuba, Jennifer and Puljak, Ema and Pierini, Maurizio and Wozniak, Kinga Anna",
  title     = "{Unsupervised New Physics Detection at 40 MHz: Training Dataset}",
  version   = "v2",
  publisher = "Zenodo",
  doi       = "10.5281/zenodo.5046428",
  url       = "https://doi.org/10.5281/zenodo.5046428",
  year      = "2021"
}

@dataset{zenodo:lq,
  author        = "Aarrestad, T. and others",
  title         = "{Unsupervised New Physics detection at 40 MHz: $LQ \to b\tau$ Signal Benchmark Dataset}",
  doi           = "10.5281/zenodo.7152599",
  url           = "https://doi.org/10.5281/zenodo.7152599",
  publisher     = "Zenodo",
  year          = "2021"
}

@dataset{zenodo:a4l,
  author        = "Aarrestad, T. and others",
  title         = "{Unsupervised New Physics detection at 40 MHz: $A \to 4\ell$ Signal Benchmark Dataset}",
  doi           = "10.5281/zenodo.5046445",
  url           = "https://doi.org/10.5281/zenodo.5046445",
  publisher     = "Zenodo",
  year          = "2021"
}

@dataset{zenodo:h0tt,
  author        = "Aarrestad, T. and others",
  title         = "{Unsupervised New Physics detection at 40 MHz: $h^0 \to \tau\tau$ Signal Benchmark Dataset}",
  doi           = "10.5281/zenodo.5061632",
  url           = "https://doi.org/10.5281/zenodo.5061632",
  publisher     = "Zenodo",
  year          = "2021"
}

@dataset{zenodo:hctaunu,
  author        = "Aarrestad, T. and others",
  title         = "{Unsupervised New Physics detection at 40 MHz: $h^\pm \to \tau\nu$ Signal Benchmark Dataset}",
  doi           = "10.5281/zenodo.5061687",
  url           = "https://doi.org/10.5281/zenodo.5061687",
  publisher     = "Zenodo",
  year          = "2021"
}

@article{Tsoi:2024ypg,
    author = "Tsoi, Ho Fung and Loncar, Vladimir and Dasu, Sridhara and Harris, Philip",
    title = "{SymbolNet: neural symbolic regression with adaptive dynamic pruning for compression}",
    eprint = "2401.09949",
    archivePrefix = "arXiv",
    primaryClass = "cs.LG",
    doi = "10.1088/2632-2153/adaad8",
    journal = "Mach. Learn. Sci. Tech.",
    volume = "6",
    number = "1",
    pages = "015021",
    year = "2025"
}

@article{Meurer:2017yhf,
    author = "Meurer, Aaron and others",
    title = "{SymPy: symbolic computing in Python}",
    doi = "10.7717/peerj-cs.103",
    journal = "PeerJ Comput. Sci.",
    volume = "3",
    pages = "e103",
    year = "2017"
}

@article{liu2020dynamic,
  author        = "Liu, Junjie and Xu, Zhe and Shi, Runbin and Cheung, Ray C. C. and So, Hayden K. H.",
  title         = "{Dynamic Sparse Training: Find Efficient Sparse Network From Scratch With Trainable Masked Layers}",
  eprint        = "2005.06870",
  archivePrefix = "arXiv",
  primaryClass  = "cs.LG",
  year          = "2020"
}

@article{companion,
  author        = "Abson, Jaiman and Canelli, Florencia and Guglielmi, Valentina and Niemi, Roope Oskari and Pierini, Maurizio and Sun, Chang and Vaselli, Francesco",
  title         = "{Knowledge Distillation of a Normalising Flow for Real-Time Anomaly Detection at LHC Level-1 Trigger}",
  eprint        = "2609.15295",
  archivePrefix = "arXiv",
  primaryClass  = "hep-ex",
  year          = "2026"
}

@article{Tahseen:2026dnf,
  author        = "Tahseen, Tara P. A. and Clarke Hall, Noah and Konstantinidis, Nikolaos and Mart{\'\i}nez Su{\'a}rez, Paula",
  title         = "{Distilling Normalizing Flows for Real-Time Anomaly Detection at the LHC}",
  eprint        = "2608.19912",
  archivePrefix = "arXiv",
  primaryClass  = "hep-ex",
  year          = "2026"
}

@article{Schmidt:2009science,
  author        = "Schmidt, Michael and Lipson, Hod",
  title         = "{Distilling Free-Form Natural Laws from Experimental Data}",
  doi           = "10.1126/science.1165893",
  journal       = "Science",
  volume        = "324",
  number        = "5923",
  pages         = "81--85",
  year          = "2009"
}

@inproceedings{Cranmer:2020wew,
  author        = "Cranmer, Miles and Sanchez-Gonzalez, Alvaro and Battaglia, Peter and Xu, Rui and Cranmer, Kyle and Spergel, David and Ho, Shirley",
  title         = "{Discovering Symbolic Models from Deep Learning with Inductive Biases}",
  booktitle     = "{Advances in Neural Information Processing Systems 33 (NeurIPS 2020)}",
  eprint        = "2006.11287",
  archivePrefix = "arXiv",
  primaryClass  = "cs.LG",
  year          = "2020"
}

@article{Duarte:2018ite,
  author        = "Duarte, Javier and others",
  title         = "{Fast inference of deep neural networks in FPGAs for particle physics}",
  eprint        = "1804.06913",
  archivePrefix = "arXiv",
  primaryClass  = "physics.ins-det",
  doi           = "10.1088/1748-0221/13/07/P07027",
  journal       = "JINST",
  volume        = "13",
  number        = "07",
  pages         = "P07027",
  year          = "2018"
}

@article{Schulte:2025mai,
  author        = "Schulte, Jan-Frederik and others",
  title         = "{hls4ml: A Flexible, Open-Source Platform for Deep Learning Acceleration on Reconfigurable Hardware}",
  eprint        = "2512.01463",
  archivePrefix = "arXiv",
  primaryClass  = "cs.AR",
  year          = "2025"
}

@article{Tsoi:2023isc,
  author        = "Tsoi, Ho Fung and Pol, Adrian Alan and Loncar, Vladimir and Govorkova, Ekaterina and Cranmer, Miles and Dasu, Sridhara and Elmer, Peter and Harris, Philip and Ojalvo, Isobel and Pierini, Maurizio",
  title         = "{Symbolic Regression on FPGAs for Fast Machine Learning Inference}",
  eprint        = "2305.04099",
  archivePrefix = "arXiv",
  primaryClass  = "cs.LG",
  doi           = "10.1051/epjconf/202429509036",
  journal       = "EPJ Web Conf.",
  volume        = "295",
  pages         = "09036",
  year          = "2024"
}

@misc{pol2023knowledgedistillationanomalydetection,
      title={Knowledge Distillation for Anomaly Detection}, 
      author={Adrian Alan Pol and Ekaterina Govorkova and Sonja Gronroos and Nadezda Chernyavskaya and Philip Harris and Maurizio Pierini and Isobel Ojalvo and Peter Elmer},
      year={2023},
      eprint={2310.06047},
      archivePrefix={arXiv},
      primaryClass={cs.LG},
      url={https://arxiv.org/abs/2310.06047}, 
}

@article{Gerlach:2025brx,
    author = "Gerlach, Lino and Kauffman, Elliott and Mallampalli, Abhishikth",
    title = "{Evaluation of Novel Fast Machine Learning Algorithms for Knowledge-Distillation-Based Anomaly Detection at CMS}",
    eprint = "2510.15672",
    archivePrefix = "arXiv",
    primaryClass = "hep-ex",
    month = "10",
    year = "2025"
}
